\documentclass[journal=jacsat,manuscript=article]{achemso}

\usepackage[normalem]{ulem}
\usepackage[utf8]{inputenc}
\usepackage{graphicx}
\usepackage{amsmath}
\usepackage{textgreek}
\usepackage{hyperref}
\usepackage{indentfirst}
\usepackage{setspace}
\usepackage[nottoc,notlot,notlof]{tocbibind}
\usepackage{appendix}
\usepackage{float}
\usepackage{titlesec}
\usepackage[version=4]{mhchem}
\usepackage{ulem}
\usepackage{lipsum}
\usepackage{upgreek}
\title{Nonlinear dielectric epsilon near-zero hybrid nanogap antennas}

\author{Romain Tirole$^1$}
\email{romain.tirole16@imperial.ac.uk}
\affiliation{The Blackett Laboratory, Department of Physics, Imperial College London, London SW7 2BW, United Kingdom}

\author{Benjamin Tilmann$^1$}
\affiliation{Chair in Hybrid Nanosystems, Nanoinstitut München, Ludwig-Maximilians-Universität München, 80539 München, Germany}

\author{Leonardo de S. Menezes}
\affiliation{Chair in Hybrid Nanosystems, Nanoinstitut München, Ludwig-Maximilians-Universität München, 80539 München, Germany}
\alsoaffiliation{Departmento de Física, Universidade Federal de Pernambuco, Recife-PE 50670-901, Brazil}

\author{Stefano Vezzoli}
\affiliation{The Blackett Laboratory, Department of Physics, Imperial College London, London SW7 2BW, United Kingdom}

\author{Riccardo Sapienza}
\email{r.sapienza@imperial.ac.uk}
\affiliation{The Blackett Laboratory, Department of Physics, Imperial College London, London SW7 2BW, United Kingdom}

\author{Stefan A. Maier}
\affiliation{School of Physics and Astronomy, Monash University, Clayton, Victoria 3800, Australia}
\alsoaffiliation{The Blackett Laboratory, Department of Physics, Imperial College London, London SW7 2BW, United Kingdom}

\begin{document}
$^1$ The authors contributed equally to this work.
\begin{abstract}

High-index Mie-resonant dielectric nanostructures provide a new framework to manipulate light at the nanoscale. In particular their local field confinement together with their inherently low losses at frequencies below their band-gap energy allows to efficiently boost and control linear and nonlinear optical processes. Here, we investigate nanoantennas composed of a thin indium-tin oxide layer in the center of a dielectric Gallium Phosphide nanodisk. While the linear response is similar to that of a pure GaP nanodisk, we show that the second and third-harmonic signals of the nanogap antenna are boosted at resonance. Linear and nonlinear finite-difference time-domain simulations show that the high refractive index contrast leads to strong field confinement inside the antenna's ITO layer. Measurement of ITO and GaP nonlinear susceptibilities deliver insight on how to engineer nonlinear nanogap antennas for higher efficiencies for future nanoscale devices.

\end{abstract}

\maketitle


\section{Introduction}

The implementation of nonlinear optics at the nanoscale and at visible to near-infrared (near-IR) wavelengths is of great interest for technical applications such as quantum light sources, optical circuits and frequency control or biophotonics. Although nanoantennas provide a gateway to efficiently couple incident light into structures at the nanoscale, other aspects such as small interaction volume or low damage threshold remain challenging. In particular, the latter limits the possible usage of plasmonic structures made of (noble) metals, together with their significant Ohmic losses. Hence, all-dielectric materials have proven to be a promising alternative platform for nanophotonics. Of particular interest are group IV and III-V semiconductors that generally have large refractive indices, high nonlinear susceptibilities as well as a high transparency for light energies below their electronic band gap, making them excellent candidates for nonlinear nanoscale applications~\cite{kuznetsov_optically_2016,krasnok_nonlinear_2018,koshelev_subwavelength_2020}. Similar characteristics appear in two-dimensional materials and more specifically transition metal dichalcogenides, however, their naturally small interaction volume yields only low absolute signal and lowers the damage threshold due to high excitation intensities~\cite{you_nonlinear_2019}.

All-dielectric nanoantennas have shown efficient harmonic generation~\cite{miroshnichenko_nonradiating_2015,grinblat_enhanced_2016,tilmann_nanostructured_2020,cambiasso_bridging_2017,carletti_enhanced_2015,carletti_shaping_2016,sanatinia_modal_2014,camacho-morales_nonlinear_2016,marino_spontaneous_2019} and ultrafast nonlinear optical switching~\cite{grinblat_ultrafast_2019,shcherbakov_ultrafast_2015,shcherbakov_ultrafast_2017,della_valle_nonlinear_2017} in various materials such as Silicon, (Aluminum) Gallium Arsenide, Gallium Phosphide or Germanium. By acting as nanocavities they open the path to larger interaction volumes while maintaining strong resonant behavior with a high tolerance to incident field intensities. Particularly the magnetic dipole~\cite{cambiasso_bridging_2017,carletti_enhanced_2015,marino_spontaneous_2019,tilmann_nanostructured_2020} and anapole resonances~\cite{grinblat_enhanced_2016,miroshnichenko_nonradiating_2015} of nanodisks have  been exploited for harmonic generation given their characteristic strong field confinement within the medium. These single-particle modes can also be used as building blocks for collective resonances: dielectric metasurfaces with various geometries were shown to exhibit high quality factors~\cite{vabishchevich_enhanced_2018,liu_all-dielectric_2018,campione_broken_2016,shcherbakov_photon_2019}, emerging from bound states in the continuum or constructive interference between single-antenna modes and lattice resonances. 

In addition to mode engineering by changing the geometry of nanostructures, an alternative approach is based on the integration of multiple materials in a single antenna~\cite{zhang_ideal_2019,guo_large_2016,aouani_third-harmonic-upconversion_2014,suresh_enhanced_2021,maccaferri_enhanced_2021}. In particular, creating a gap inside a high-index dielectric structure in the lateral direction, filled either with air or a low-index material has the potential to have great impact on the linear and nonlinear properties of the system~\cite{yang_anapole-assisted_2018}. Given the optical contrast between dielectric $n_d$ and gap $n_s$, the field intensity scales as $n_d^4/n_s^4$ from the continuity of the electric displacement field at an interface, where $n_d,n_s$ are the respective refractive indices \cite{robinson_ultrasmall_2005}. Moreover, as long as the slot's dimensions are significantly smaller than that of the the effective wavelength, the far field optical properties of the antenna are retained as well as its resonant behavior, as determined by the effective refractive index. This makes these nanogap antennas promising candidates for nonlinear nanophotonic applications~\cite{patri_hybrid_2022}, however, the experimental realization of different materials in a single nanostructure remains challenging. Only recently, a nanogap using a low index material, Indium Tin Oxide (ITO), slotted in silicon was demonstrated and used for enhanced harmonic generation~\cite{thouin_broadband_2022}. ITO shows a zero-crossing of the real part of the dielectric function in the near-infrared (near-IR), allowing for the realisation of a nanogap with high optical contrast. It was also shown that ITO exhibits large optical nonlinearities in the vicinity of this so-called epsilon-near-zero (ENZ) regime~\cite{alam_large_2016,carnemolla_degenerate_2018,luk_enhanced_2015}.

In this work, we realize and characterize a nanogap antenna made of high-refractive index Gallium Phosphide (GaP) and a nanolayer of ITO and compare it to a pure GaP antenna of similar dimensions. We design the nanostructures to exhibit strong optical resonances in the ENZ region, dominated by magnetic dipole (MD) and electric (ED) contributions. Under illumination with near-IR light, we observe strong second- and third-harmonic generation (SHG, THG), whereas the nanogap antenna exceeds the efficiency of the GaP-only case particularly for excitation at the ED resonance. We show that this enhancement appears over a broad wavelength range, exceeding the ENZ regime of ITO. To explain this result, we perform a thorough analysis of the nonlinear coefficients of both materials and conduct nonlinear scattering simulations. Our results proposes the concept of nanogap antennas as a new and flexible platform with a wide range of possible applications in nonlinear optics. Furthermore, it opens the path of integrating ITO or other ENZ materials into dielectric structures to enhance specific optical resonances of the  directivity of emission, as well as time-modulated antennas for frequency control and holography \cite{karl_frequency_2020,karimi_time-varying_2023}.


\section{Design and fabrication}

\begin{figure}[h!]
    \centering
    \includegraphics[width=\textwidth]{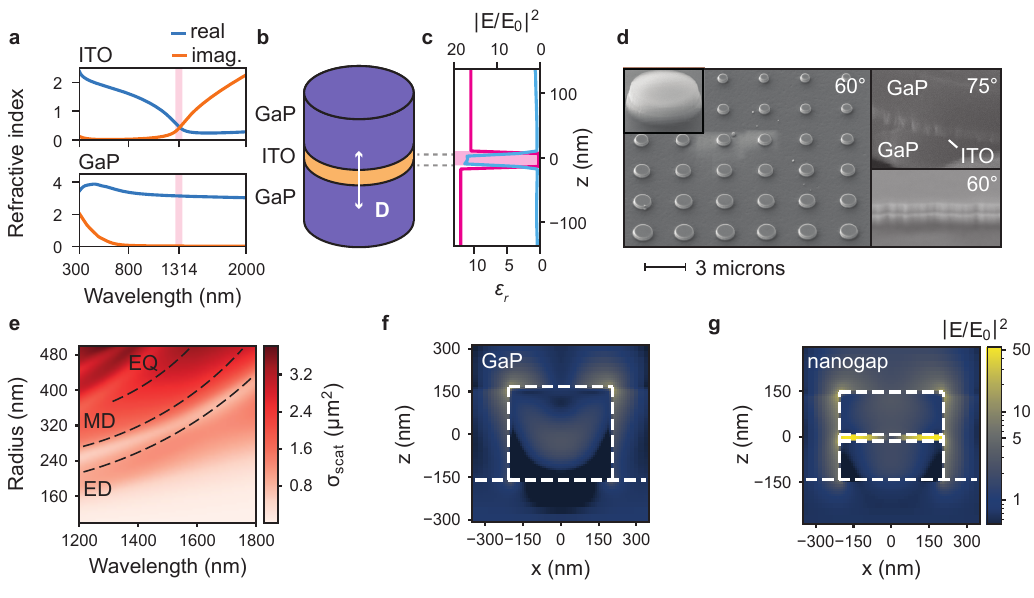}
    \caption{GaP and nanogap antennas for nonlinear light generation. \textbf{a} Refractive index spectrum of ITO: real (blue curve) and imaginary (orange curve) parts. The ENZ region is highlighted with the red-shaded area. \textbf{b} Sketch of the nanogap antenna concept. The displacement field is represented by the white double-ended arrow. \textbf{c} Diagram of the expected field intensity distribution (pink) as a function of the centered antenna depth, resulting from the permittivity contrast (light blue). \textbf{d} Scanning electron microscopy images of the sample, taken at angles of 60 and 75 degrees (tilted).  \textbf{e} FDTD simulation of the scattering cross section of a GaP antenna as a function of pump wavelength and antenna radius. The ED, MD and EQ modes are highlighted with the black dashed curves. \textbf{f,g} Simulated field intensity profile in logarithmic scale, the XZ plane for a radius of 200 nm and 1160 nm wavelength for the GaP (f) and nanogap (g) antennas. White dashed lines show the contour of the different materials.   }
    \label{fig:fig1}
\end{figure}

GaP-ITO-GaP films are fabricated by a multi-step sputtering process at elevated temperature, more details can be found in the Supplementary materials. The materials where chosen such that a high optical contrast between the two layers is achieved, confirmed by the measured refractive indices shown in Figure \ref{fig:fig1}a, while maintaining losses low at $\omega$ and $2\omega$. The ENZ point of the ITO layer lies around 1300~nm, a regime where GaP shows a high refractive index and virtually no optical losses. In a nanodisc with an ITO film located in the center, as sketched in Fig. \ref{fig:fig1}b, the contrast of the permittivity at the ENZ wavelength reaches a maximum of more than ten, which leads to a strong confinement of the electric field inside the ITO layer (see Fig. \ref{fig:fig1}c. The fabrication of the nanostructures is carried out by electron-beam lithography (EBL) and subsequent reactive ion etching (RIE), more details can be found in the Supplementary Information. The distance between individual antennas is kept at $3\,\upmu\text{m}$ to prevent cross-antenna coupling. GaP-only nanodisks with identical dimensions were fabricated following the same process. A scanning electron microscopy (SEM) image of the nanogap discs with varying radii is shown in Fig. \ref{fig:fig1}d (left). Looking at the tilted SEM image in the right hand side of Fig. \ref{fig:fig1}d, one can clearly observe the three distinct layers which confirms the successful fabrication of the nanogap antennas. 

We perform finite-difference time-domain (FDTD) simulations to investigate the scattering properties of the fabricated structures. Fig. \ref{fig:fig1}e shows the evaluated scattering cross section for GaP-only structures, in dependence of the excitation wavelength and nanodisc radius. Clearly, a set of distinct resonances can be identified. By comparison with a multipole decomposition shown in the Supplementary Information, three main contributions can be identified as indicated in Fig. \ref{fig:fig1}e. At larger radii, higher order as the electric quadrupole (EQ) dominate the scattering spectrum, whereas smaller radii are dominated by MD and ED. A similar simulation carried out for the hybrid nanodiscs is shown in Supplementary Information and confirms that the far-field response is only marginally affected by the presence of the ITO layer (which makes only 7.5\% of the volume of the antenna). However, the continuity of the displacement field vector $\textbf{D}$ requires a change in the electric near-field distribution and consequently strongly confines the electric near-field into the low-index region (pink-shaded area in Figure \ref{fig:fig1}c). This is even more pronounced when comparing the simulated near field distributions of the GaP-only nanodisc in Fig. \ref{fig:fig1}f and of the nanogap antenna in Fig. \ref{fig:fig1}g, for a wavelength of 1160 nm and an antenna with a radius of 200 nm. For the latter, the field is clearly highly concentrated within the 20 nm thin ITO layer with distribution towards the edges of the disc. Contrarily, the near field of a GaP-only antenna under similar conditions has a field broadly confined over the antenna's volume, with maximum intensity values one order of magnitude smaller.\\



\section{Harmonic measurements}

\begin{figure}[h]
    \centering
    \includegraphics[width=\textwidth]{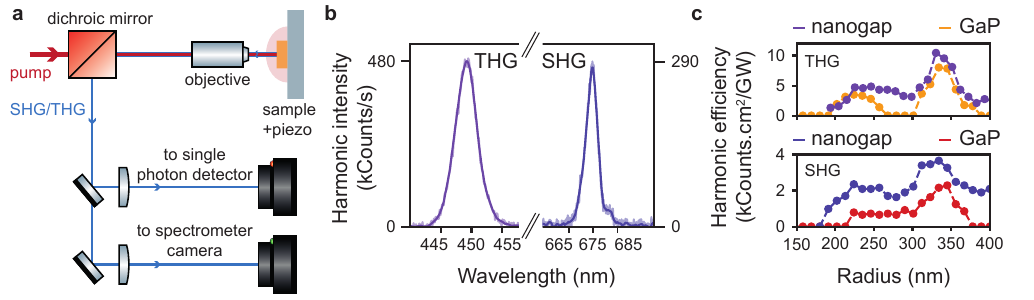}
    \caption{Measurement of harmonic generated light from nanodisks. \textbf{a} Experimental microscopy setup for the spectral characterisation of antennas. The reflected harmonic from the sample can be sent for detection either by a single photon detector or CCD camera spectrometer. \textbf{b} Second and third harmonic spectra from a nanogap antenna of radius of 330 nm and illumination wavelength of 1350 nm. \textbf{c} Comparison of THG (top) and SHG (bottom) efficiency dependence on radius between the nanogap (purple and blue curves) and the GaP-only (yellow and red curves) nanodisks, for a wavelength of 1350 nm.}
    \label{fig:fig2}
\end{figure}

The GaP and nanogap discs are investigated with respect to their nonlinear optical response by using a nonlinear microscopy setup, based on a tunable wavelength laser with 100 kHz repetition rate and 225 fs (FWHM) pulses. The generated nonlinear signal is collected in reflection geometry (see Fig. \ref{fig:fig2}a) and then either sent to a single photon detector or a spectrometer to respectively map the resonances as a function of radius and illuminating wavelength, or measure the light's spectrum. To calculate the intensity at each wavelength, average excitation power and beam size are measured at the focal position. More details of the experimental setup can be found in the Supplementary Information. 

For an excitation wavelength of $\lambda_0 = 1350$ nm and a nanogap disc radius of $\sim$ 330 nm, the corresponding SHG and THG spectrum are shown in Fig. \ref{fig:fig2}b. The collected light is indeed harmonic centered at $\lambda_{SHG} = 675\,\text{nm}=1350\,\text{nm}/2$ and $\lambda_{THG} = 450\,\text{nm}=1350\,\text{nm}/3$, with no evidence for photoluminescence. Furthermore, the nonlinear character of the observed signal is confirmed by power dependence measurements (see Supplementary Information). Measurements on a GaP-only disc yield similar results, however, from the amorphous nature of the GaP it is expected that the SHG is largely generated by surface instead of bulk contributions \cite{tilmann_comparison_2023}.


Fig. \ref{fig:fig2}c shows the SHG and THG signal at a fixed excitation wavelength of 1350 nm in dependence of radius for GaP-only and GaP-ITO antennas. In the case of THG (upper panel), the enhancement of the nanogap (purple) against GaP (dark yellow) is limited at the visible ED and MD modes, with only an enhancement by a factor of 1.3 (respective values of 10.45 and 8.01 kHz.cm$^2$/GW). Yet, the resonances are significantly broadened over a range of radii, with THG counts staying at a constant level between 230 and 300 nm radius. For SHG on the other hand, the nonlinear signal is enhanced over a broad range of radii (blue against red line, lower panel). This can be explained by the added amount of interfaces between layers, essential in the generation of the second harmonic as it increases the surface area where SHG occurs \cite{shen_optical_1989}, along with a potential higher second order nonlinear susceptibility from the surface of ITO.

\begin{figure}[h]
    \centering
    \includegraphics[width=\textwidth]{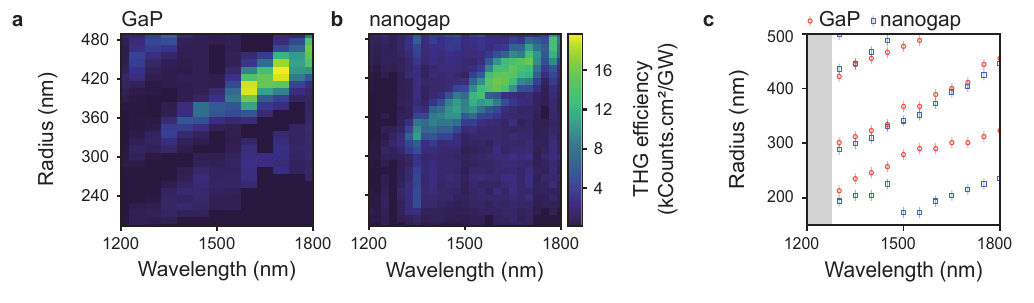}
    \caption{Characterisation of the nanodisk's nonlinear signals. \textbf{a,b} Experimentally measured THG signal in counts per second per unit intensity as a function of disk radius and pump wavelength for GaP (a) and nanogap (b) disks. \textbf{c} Extracted resonant wavelength for the ED, MD and EQ modes as a function of radius for the nanogap antenna (blue diamonds), GaP antenna (red diamonds)}
    \label{fig:fig3}
\end{figure}

Next, we characterise the spectral dependence of the harmonic generation and the evolution of the nanodisc resonances by varying the incident light pulse wavelength. Fig. \ref{fig:fig3}a shows the measured THG counts per unit intensity for antennas without the ITO layer, in dependence of radius and excitation wavelength. Three distinct resonances are clearly visible, and can be associated to the scattering simulation in Figure \ref{fig:fig1}e. It follows that the lower resonances is caused by the scattering peak that was associated with the ED resonances, whereas the second follows the course of the MD scattering dip. The resonance that is visible at the largest radii can be assign to the higher order electric quadrupole (EQ). The THG signal that is generated from the MD is more than one order of magnitude larger than for ED and EQ. This can be explained by near field simulations (see Supplementary Material), where the MD strongly concentrates the electric field inside the volume of the nanostructures. Furthermore, this is confirmed by nonlinear scattering simulations that are shown in the Supplementary Material and which qualitatively reproduce the measured THG signal with good agreement.

Following the same procedure, THG was measured for the hybrid GaP-ITO nanostructures, with the spectral and radial dependence shown in Figure \ref{fig:fig3}b. Again, the previous three resonances can clearly be seen, with the ED, MD and EQ appearing at similar radii and wavelengths as for the GaP-only structures. One can see the MD mode yields less signal further in the IR, around 1650 nm wavelength, but presents a higher generation efficiency in range near its ENZ wavelength (1300-1400 nm). Furthermore, the mode is spectrally broader, and the nanogap antenna exhibits a higher THG efficiency off resonance compared to the GaP-only case.

The nanogap antenna shows a distinctive feature in mode evolution when compared to the GaP antenna in its ED resonance. Fig. \ref{fig:fig3}c represents the extracted resonant radii of the disks for each excitation wavelength. Each diamond corresponds to a mode's most efficient radius for THG for a given wavelength. All three modes are clearly visible for GaP (light red), and as expected the optimal antenna resonant radius increases with wavelength for each mode as the GaP antenna is a cavity. This is verified by nonlinear scattering simulations (see Supporting Information), which agrees very well with the experimental measurement. For the GaP-ITO antenna, a similar behavior can be observed for the MD and EQ resonances, agreeing well with the course of the GaP-only case and the simulations. For the ED however, a particular dip in the resonant wavelength can be observed for the ED mode, that happens around the ENZ region. Here, the resonant radius drops by $\sim$ 50 nm when the illuminating pump wavelength is increased by the same amount.



\section{Nonlinear susceptibility and field distribution}

\begin{figure}[h]
    \centering
    \includegraphics[width=\textwidth]{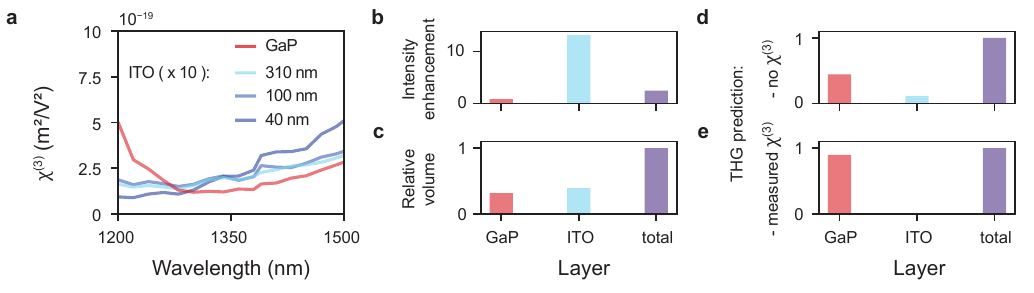}
    \caption{Mechanisms reducing harmonic generation from the ITO layer. \textbf{a} Measured $\chi^{(3)}$ of various ITO thin films, with 40 nm (dark blue curve), 100 nm (blue curve) and 310 nm (light blue curve) thicknesses, and of GaP (red curve). The nonlinear susceptibilities of ITO are multiplied by 10 for comparison. \textbf{b} Average intensity enhancement with regards to the incoming beam, in the ITO (light blue bar) and GaP (red bar), layers as well as over the entire antenna (purple bar), simulated using FDTD for a 320 nm nanodisk at the MD resonant frequency of 1285 nm. \textbf{c} Corresponding intensity enhancement integrated normalised by the volume of the layers or antenna: the comparatively small size of the ITO layer reduces its contribution to the THG signal. \textbf{d} THG prediction using FDTD, setting $\chi^{(3)}_{\textrm{GaP}}=\chi^{(3)}_{\textrm{ITO}}$: a poor overlap with the harmonic field profile reduces the ITO layer's contribution even further in comparison to the GaP layers. Note that constructive interference between the ITO and GaP layers enhances the harmonic signal. \textbf{e} THG prediction using FDTD and the experimentally measured nonlinear susceptibility of the materials: the ITO contribution to THG is greatly reduced in comparison to the average intensity enhancement presented in panel b.}
    \label{fig:fig4}
\end{figure}

Although an enhancement of the nonlinear signal of a factor up to 2 at resonance from the nanogap compared to the GaP-only antennas could be observed, the effects is not comparable to 100$\times$ field enhancement reported in \cite{thouin_broadband_2022}. To address this discrepancy, we measure and compare the $\chi^{(3)}$ of GaP and ITO in the following. This is done by measuring the samples under weakly focused illumination after which the THG light is collected. Then, the measured harmonic power is processed and compared to the illuminating power to extract a value of the nonlinear susceptibility (see Supplementary Information). A 400 nm thick GaP thin film as well as three thin films of ITO with various thicknesses (40, 100 and 310 nm) are investigated. Note that an additional 20 nm ITO thin film was measured, but yielded too weak harmonic signal in the present configuration to give an accurate value for the nonlinear susceptibility. The resulting third harmonic nonlinear susceptibility spectra are presented in Fig. \ref{fig:fig4}a. As can be seen, GaP (in red) has a $\chi^{(3)}$ an order of magnitude above that of ITO (in blue), with a value of 6.95 $\times 10^{-19}$ m$^2$/V$^2$ at $1350\,\text{nm}$ wavelength. All three ITO samples, with various thicknesses and different ENZ wavelengths in the near-infrared region, have comparable third harmonic susceptibilities. For this reason, we assume the nonlinear susceptibility of the $21\,\text{nm}$ layer of ITO within the nanogap antennas to be equal to that of the $40\,\text{nm}$ layer. The reason that the measured $\chi^{(3)}$ is lower than previous reports using Z-scan or pump-probe spectroscopy \cite{alam_large_2018,carnemolla_degenerate_2018} is the following. These techniques are efficient at estimating third order nonlinearities for the Kerr effect, corresponding to a nonlinear susceptibility tensor of taking the form of $\chi^{(3)}(\omega,\omega,\omega,-\omega)$ where the first frequency refers to the outgoing wave and the following ones to the waves interacting via the nonlinearity. For THG, the nonlinear susceptibility is evaluated for different frequencies of $\chi^{(3)}(3\omega,\omega,\omega,\omega)$ and can thus take a different value. Thus, observing strong Kerr nonlinearities in ITO does not guarantee observing efficient THG.

We further investigate the contributions of the ITO and GaP layers within the antenna by looking at FDTD simulations, more precisely at the average field enhancement (normalised to the illuminating field intensity), shown in Fig. \ref{fig:fig4}b for a radius of 320 nm, at the MD resonant wavelength of 1285 nm. Clearly, the field is highly concentrated into the ITO layer (light blue box), with a predicted average enhancement of 13.17. From this, the contribution of the ITO layer to the generated THG signal could be approximated to be $13.17^3$ larger than that of the GaP layers. But evidently, this did not lead to an increase of the total THG signal in the experiment, which we explain by three contributing factors.


First, the ITO layers occupies only 7.5\% of the total. This explains the low shift in the overall antenna resonant radius, but as Fig. \ref{fig:fig4}c shows it also plays against the third harmonic enhancement from the nanogap effect. Multiplying the average field enhancement by the relative volume logically reduces the contribution of the ITO to the total THG by a factor of 10, on a now similar scale to that of GaP. 

Second, the antenna's resonance at the harmonic wavelength plays an important role in the harmonic generation process \cite{koshelev_subwavelength_2020,obrien_predicting_2015}. In order to enhance the generated signal, the nanodisk needs to provide good overlap between the resonances at the fundamental and the harmonic. Nonlinear FDTD simulations (see Supplementary Information) in Fig. \ref{fig:fig4}d shows this is not the case in the ITO layer, which suffers from this constraint in comparison to the GaP layers. Nevertheless, constructive interference between the THG signal generated by the ITO and GaP layers can enhance the resulting harmonic signal.

Finally, with the additional effect of the nonlinear susceptibility, the contribution of the ITO layer to the total harmonic generated signal then becomes modest, as shown in Fig. \ref{fig:fig4}e.

We have thus shown that the insertion of an ITO nanogap in a high-index dielectric antenna brings little advantage in third harmonic enhancement if the base material shows high nonlinearities. Similar results are expected to hold for Al-GaAs, Ge, or \ce{LiNbO3}, with ITO or AZO as a nanogap layer. Second harmonic generation, on the other hand, is enhanced, which we attribute to the fact it is not generated in the bulk of centrosymmetric materials such as a-GaP or ITO \cite{boyd_nonlinear_2020,shen_optical_1989}, and is proportional to area and not volume, thus mitigating the limited height of ITO. Furthermore, the field distribution is enhanced at the surface of the ITO, while for GaP the field is more evenly distributed throughout the volume.

Instead, we suggest an alternative application for nonlinear nanogap antennas. To take advantage of the strong Kerr effect and field enhancement within the ITO layer, the nanogap antennas could be used for ultrafast switching and frequency modulation via time-varying effects. By experiencing a strong and swift change in permittivity under pumping at the resonant wavelength, a probe beam at the same wavelength should undergo changes in carrier frequency \cite{pang_adiabatic_2021} and spectral width \cite{tirole_saturable_2022}. Then, factors such as overlap and interference of the harmonic field are ruled out, while making use of the strong Kerr nonlinearity of ITO.


\section{Conclusion}

In summary, we have characterised the crossing between the resonances of a GaP high-index dielectric antenna and the nanogap effect from an ITO thin film, by measuring a broadband enhancement of SHG and THG signal from the nanodiscs. The comparison to a GaP-only antenna shows the conservation of the mode resonances and a significantly increased second harmonic signal across all nanogap antenna radii, while the third harmonic, though slightly enhanced, shows a limited increase in efficiency. We attribute this to a combination of lowered parametric nonlinearities in ITO in comparison to GaP, as well as the limitations of volume and field overlap affecting the overall signal generated by the nanogap layer. This demonstrates that engineered antennas with ENZ nanolayers can be used for nonlinear light generation. Further development of these nanodisks could leverage the exceptional properties of ITO and other transparent conducting oxides for ultrafast switching or time-varying effects \cite{alam_large_2016,galiffi_photonics_2022}.


\begin{acknowledgement}

The authors acknowledge funding from the Engineering and Physical Sciences Research Council (EP/V048880). S.A.M. additionally acknowledges the Australian Research Council and the Lee-Lucas Chair in Physics.

\end{acknowledgement}


\begin{suppinfo}

Supporting Information is available from the Wiley Online Library or from the author.

\end{suppinfo}


\bibliography{main}

\end{document}



\section{Sample Fabrication}

The GaP and ITO films were fabricated by subsequent sputter deposition using an Angstrom deposition tool. During the procedure, the sample holder was kept at a constant temperature of 350°C, while it was allowed to cool down between the different deposition steps to avoid mixing of the different layers. As substrate, borosilicate cover slips were used and previously cleaned with acetone and isoprpyl alcohol (IPA) and ultrasonication. Finally, a 70~nm layer of silicon dioxide was deposited on top of the layer, using plasma enhanced chemical vapour deposition (PECVD, Oxford Instruments). The EBL was preformed using Poly(methyl methacrylate) (PMMA) as photoresist and carrried out with a 20~\textmu m aperture, 30~kV acceleration voltage and a dose factor of 400 \textmu C/cm$^2$ (e-LiNE, Raith GmbH). After development in methyl isobutyl ketone:IPA (ratio 1:3) for about 50~s and 10~s of oxygen plasma cleaning (20~sccm, 40\% power), a 50~nm gold film was deposited on top using electron beam evaporation at ultrahigh vacuum. Next, the lift-off process was performed using Remover 1165 (...) at 80°C for several hours, removing the remaining PMMA and leaving only the desired structures made of gold. Finally, several steps of inductively-coupled plasma reactive-ion etching (ICP-RIE) were performed to transfer the design into the nanogap material. First, florine based chemistry was used to etch into the silicon dioxide layer, while clorine based chemistry removes the nanogap material. In between the two steps, gold etchant (Aldrich) was used to remove the gold etchant.

\section{Optical Setup}

220~fs optical pulses with a 100~kHz repetition rate are generated with a LightConversion Pharos laser, and wavelength is controlled via a LightConversion Orpheus OPA. Light is sent through a 60X objective with NA 0.85, and collected in reflection. Harmonic signal is separated from the pump beam using a dichroic beam splitter, and second-harmonic-generation is filtered out using a 750~nm short-pass filter. THG is sent either to a Micro Photon Devices PDM single photon detector, or to a Princeton Instruments spectrometer with a Pixis CCD camera. Precise control of the illumination area is achieved with a Piezoconcept C3 3-axis piezo stage.

\section{Power dependence measurements}

The measured power dependences are consistently lower than expected from Nonlinear Optical theory, yet second and third harmonic spectra confirm the nature of the harmonics and do not show the presence of other signals. We suggest this is could be due to the antennas being in a saturation regime near the damage threshold, preventing further generation.

\begin{figure}[h]
    \centering
    \includegraphics[width=\linewidth]{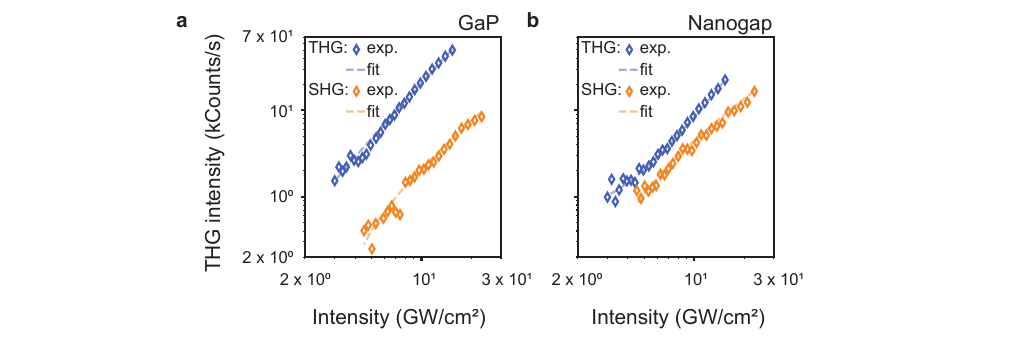}
    \caption{Power dependence of the harmonic generated signal from the antennas. \textbf{a,b} Intensity dependence of the second (orange) and third (blue) harmonic signals in (a) GaP and (b) the nanogap antenna. The fitted curves are shown with dashed lines.}
    \label{fig:suppfig_powdep}
\end{figure}

\begin{center}
    \begin{tabular}{|c|c|c|}
        \hline
        Antenna & Fitted SHG slope & Fitted THG slope \\
        \hline
        GaP & 1.70 & 2.11 \\
        Nanogap & 1.45 & 2.42 \\
        \hline
    \end{tabular}
\end{center}

\section{Nonlinear Scattering Theory}

Our nonlinear scattering model is implemented following O'Brien et al.'s~\cite{obrien_predicting_2015}. Using the Lorentz reciprocity of the medium, for current sources $\boldsymbol{j}_1(\boldsymbol{r},\omega)$ and $\boldsymbol{j}_2(\boldsymbol{r},\omega)$ generating fields $\boldsymbol{E}_1(\boldsymbol{r},\omega)$ and $\boldsymbol{E}_2(\boldsymbol{r},\omega)$,

$$\int\boldsymbol{j}_2(\boldsymbol{r},\omega)\cdot\boldsymbol{E}_1(\boldsymbol{r},\omega)dV=\int\boldsymbol{j}_1(\boldsymbol{r},\omega)\cdot\boldsymbol{E}_2(\boldsymbol{r},\omega)dV$$

\begin{figure}[h!]
    \centering
    \includegraphics[width=\linewidth]{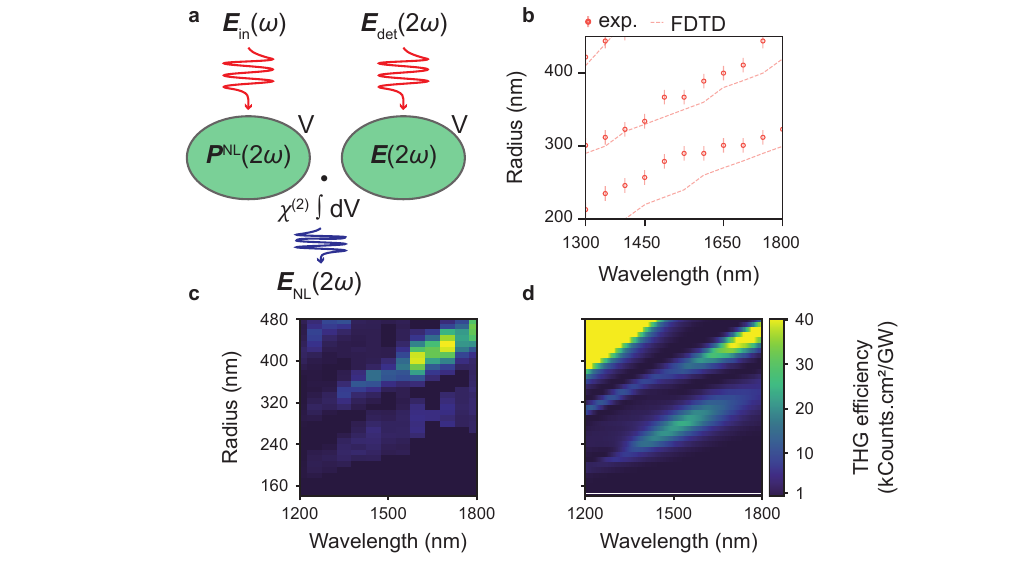}
    \caption{Nonlinear scattering theory. \textbf{a} Diagram of nonlinear scattering theory. The electric field distribution in the structure is simulated via FDTD at both the fundamental and harmonic level, the polarisation and overlap computation gives the predicted harmonic generated signal. \textbf{b} Comparison of the resonant wavelengths for harmonic generation for each antenna radius between experiment (circles) and nonlinear scattering theory (dashed curves) for a GaP antenna. \textbf{c,d} Experimentally measured (c) and predicted (d) THG signal in counts per second per unit intensity as a function of disk radius and pump wavelength for a GaP antenna.}
    \label{fig:suppfig_nlscat}
\end{figure}

Reciprocity implies the using the harmonic light generated by the currents induced by the nonlinear polarisation as a source, the linearly generated currents will be the same as the ones resulting from the nonlinear polarisation. Thus, computing the overlap integral between the nonlinear currents induced by the pump (computed from the linear induced currents the shape of the $\chi^{(3)}$) and the linear currents induced by a source at the harmonic wavelength in the far field, as shown in Fig. \ref{fig:suppfig_nlscat}a, will yield a THG signal as follows:

$$\boldsymbol{E}^{THG}\cdot\boldsymbol{\hat{j}}\propto i\omega^2e^{-3i\omega t}\int\boldsymbol{P}^{NL}\cdot\frac{\boldsymbol{E}}{\mid E_0\mid}dV$$

where $\boldsymbol{E}^{THG}$ is the predicted harmonic signal, $\boldsymbol{E}$ the field induced by the harmonic plane wave in the nonlinear medium and $\boldsymbol{\hat{j}}$ its polarisation vector at its source, and $\boldsymbol{P}^{NL}=\chi^{(3)}\boldsymbol{E}^{pump}$ is the nonlinear polarisation in the medium induced by the pump $\boldsymbol{E}^{pump}$. The factor of $\omega^2$ originates from the conversion of dipole to plane wave source.

We use nonlinear scattering theory to verify the consistency of our experimental results, by predicting the third harmonic generation frequency of GaP antennas, a now well understood and simulated system in literature \cite{grinblat_ultrafast_2019, tilmann_nanostructured_2020}. As can be seen in Fig. \ref{fig:suppfig_nlscat}b, the resonant wavelengths of a GaP nanodisk with a given radius are consistent between experiment and theory. Mapping of the full THG efficiency against radius and wavelength in Fig. \ref{fig:suppfig_nlscat}c,d show that nonlinear scattering theory qualitatively reproduces the experimental results. The slight overestimation of the electrical modes' efficiency (ED and EQ) in Fig. \ref{fig:suppfig_nlscat}d, when compared to Fig. \ref{fig:suppfig_nlscat}c, can be attributed to the simulation software's interpolation of the field in space.

We also use nonlinear scattering theory to normalise for intensities and coupling within the thin film when measuring the third-order nonlinear susceptibility of GaP and ITO, following the same procedure as presented in reference \cite{tilmann_comparison_2023}.

\clearpage
\newpage

\section{Effect of the photonic gap on the GaP nanodisk's scattering coefficient}

\begin{figure}[h!]
    \centering
    \includegraphics[width=\linewidth]{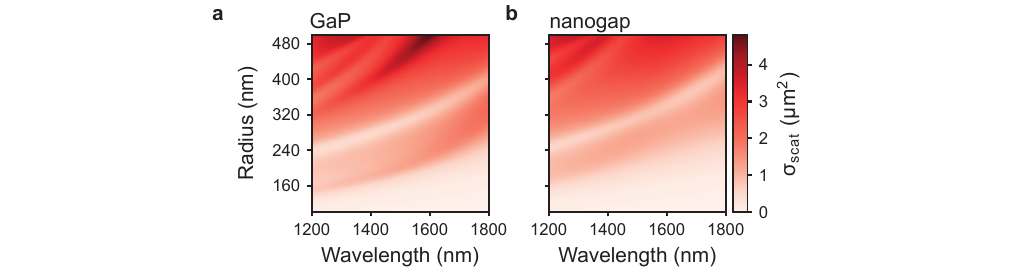}
    \caption{Scattering coefficient of nanodisks. \textbf{a,b} Scattering coefficient of (a) GaP and (b) nanogap antennas against wavelength and radius.}
    \label{fig:suppfig_scat}
\end{figure}

\section{Mode decomposition and field profile}

\begin{figure}[h!]
    \centering
    \includegraphics[width=\linewidth]{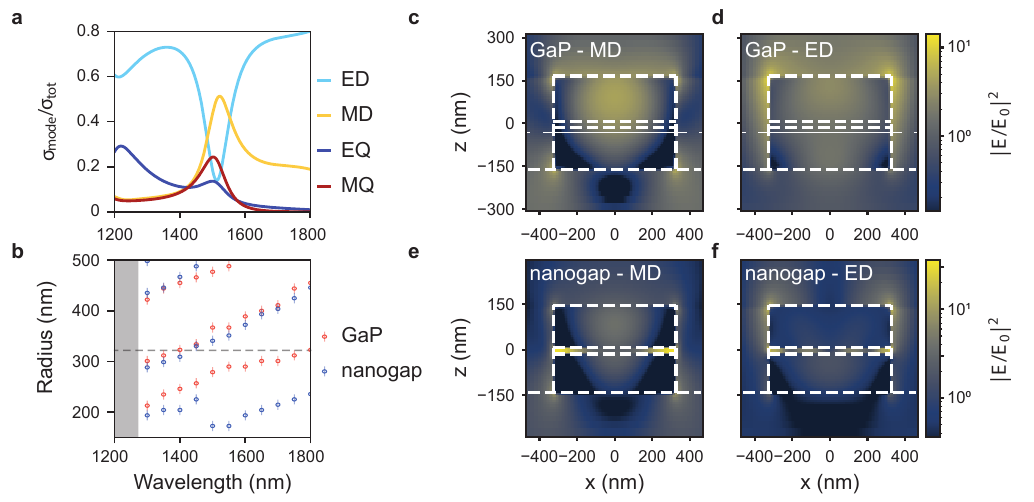}
    \caption{Mode decomposition and field distribution in the GaP and photonic GaP antennas. \textbf{a} Relative contribution of dipolar and quadrupolar modes to the scattering coefficient in a 320 nm radius GaP antenna. The MD contribution becomes strong in the vicinity of 1500 nm, where the ED is suppressed. \textbf{b} Experimentally measured resonant wavelengths for THG against radius. The grey dashed line indicates the radius of 320 nm for which the simulations in panels a and c-f are shown. \textbf{c-f} Simulated field distribution in the $(x,z)$ plane for nanodisks of radius 320 nm, on a logarithmic scale. The MD modes are plotted for wavelengths of 1450 nm and 1414 nm respectively for the GaP and nanogap antennas, where the simulation predict the peak mode intensity. The ED mode is shown for a wavelength of 1700 nm for both nanodisks as it is broader.}
    \label{fig:suppfig_mode}
\end{figure}

For the multipole decomposition, the electric and magnetic near fields are simulated and saved using Lumerical FDTD. Following, the calculations are performed using the MENP package for MATLAB (cite DOI:10.1364/osac.425189) which solve the scattering cross section in terms of the dipole and quadrupole contributions, for both the electric (ED/ EQ) and magnetic (MD/ MQ) moments as shown in Fig. \ref{fig:suppfig_mode}a for a GaP antenna with 320 nm radius.

Comparing the resulting mode decomposition to experimental data, we can expect from Fig. \ref{fig:suppfig_mode}a to observe a MD resonance in the vicinity of 1500 nm and a broad ED resonance in the 1700-1800 nm range. The measurement of the GaP antenna's resonant wavelength at that radius confirms this in Fig. \ref{fig:suppfig_mode}b, where the dashed grey line highlights the 320 nm radius where the resonances are observed.

Getting a closer look at the field distribution in the antenna, here in the $(x,z)$ plane where $z$ is the direction of propagation of light and $x$ the direction of polarisation of light, one can notice the field is more localised within the volume of the antenna for the MD in Fig \ref{fig:suppfig_mode}c than the ED mode in Fig. \ref{fig:suppfig_mode}d. For the nanogap antenna, the same difference applies between the two modes, with a lower contrast as the photonic gap concentrates the electric field in both cases. Note that for a radius of 320 nm, both the MD and ED resonance are very broad spectrally, leading to a visible superposition of the two even though the ED is suppressed in one case and enhanced in the other.

\clearpage
\newpage

\bibliography{supplement}